\documentclass[letterpaper, 10 pt, conference]{ieeeconf}  % Comment this line out if you need a4paper

\IEEEoverridecommandlockouts                              % This command is only needed if 
\usepackage[hyphens]{url}  % DO NOT CHANGE THIS
\usepackage{graphicx} % DO NOT CHANGE THIS
\usepackage{algorithm}
\usepackage{algorithmic}
\usepackage{microtype}      % microtypography
\usepackage{xcolor}         % colors
\usepackage{graphicx}
\usepackage{subfigure}
\usepackage{amsmath}
\usepackage[T1]{fontenc}

\title{\LARGE \bf
CCKS: Consensus-based Communication and Knowledge Sharing
}

\author{Jinyuan Zu$^{1,+}$, Xiaowei Lv$^{1}$, Yongcai Wang$^{1,*}$, Deying Li$^{1}$, Yunjun Han$^{2}$ \\ Wenping Chen$^{1}$, Fengyi Zhang$^{3}$ and Naiqi Wu$^{4}$% <-this % stops a space
\thanks{*This work was supported by the the National Natural Science Foundation of China Grant No.12071478 and No.61972404, Public Computing Cloud, Renmin University of China, and the Blockchain Lab. School of Information, Renmin University of China.}% <-this % stops a space
\thanks{$^{1}$School of Information, Renmin University of China, Beijing 100872, China}
\thanks{$^{2}$State Key Laboratory of Multimodal Artificial Intelligence Systems, Beijing Engineering Research Center of Intelligent Systems and Technology, Institute of Automation, Chinese Academy of Sciences, Beijing 100190, China}
\thanks{$^{3}$The Information Science Academy, China Electronics Technology Group Corporation, Beijing 100043, China}
\thanks{$^{4}$Department of Mechatronics Engineering, Guangdong University of Technology, Guangzhou 510090, China}
\thanks{$^{+}$Jinyuan Zu is the co-first author}
\thanks{$^{*}$Yongcai Wang is the corresponding author: {\tt\small ycw@ruc.edu.cn}}
}

\begin{document}

\maketitle
\thispagestyle{empty}
\pagestyle{empty}

%%%%%%%%%%%%%%%%%%%%%%%%%%%%%%%%%%%%%%%%%%%%%%%%%%%%%%%%%%%%%%%%%%%%%%%%%%%%%%%%
\begin{abstract}

In Decentralized Training and Decentralized Execution (DTDE) for cooperative Multi-Agent Reinforcement Learning (MARL), action-advising-based knowledge sharing promotes interpretable and scalable cooperation among agents. However, current action advising approaches often adhere too much to the teacher's guidance without evaluating teacher-student compatibility, which causes excessive advising, suboptimal stability, and degraded performance. 
To overcome these challenges, this paper presents a Consensus-based Communication and Knowledge Sharing (CCKS) framework, which allows agents to adopt recommendations based on consensus-derived constraints and to follow the teacher's instructions more smartly. This mechanism enables agents to balance exploration and learning from experienced teachers, improving overall performance. The key is the consensus model construction, for which we propose to employ contrastive learning to construct consensus models based on local observations in the agents' training phase. 
In action selection, agents score and choose actions based on consensus and shared knowledge. Designed as a plug-and-play solution, CCKS integrates seamlessly with existing DTDE algorithms. Experiments conducted in the Google Research Football environment and the complex StarCraft II Multi-Agent Challenge demonstrate that the integration with CCKS significantly improves cooperation efficiency, learning speed, and overall performance compared with current DTDE baselines. The code is available at \url{https://github.com/yuanxpy/CCKS}.

\end{abstract}

%%%%%%%%%%%%%%%%%%%%%%%%%%%%%%%%%%%%%%%%%%%%%%%%%%%%%%%%%%%%%%%%%%%%%%%%%%%%%%%%
\section{INTRODUCTION}

Multi-agent reinforcement learning (MARL) enables multiple agents to collaborate toward common goals by learning policies that maximize cumulative returns \cite{fan2020distributed, schmidt2022introduction, ye2020mastering}. Most MARL tasks with $n$ agents are modeled as Decentralized Partially Observable Markov Decision Processes (Dec-POMDPs) \cite{oliehoek2012decentralized}, emphasizing the agents only get the local observations.Based on Dec-POMDP, two paradigms have been recently proposed: ``Centralized Training and Decentralized Execution'' (CTDE) \cite{lowe2017multi, rashid2020monotonic, iqbal2019actor, liu2020pic} and ``Decentralized Training and Decentralized Execution'' (DTDE) \cite{tan1993multi, tampuu2017multiagent, peng2021facmac, jiang2022i2q}. DTDE offers greater adaptability in complex real-world scenarios by eliminating centralized training but suffers from training instability \cite{lowe2017multi} and suboptimal performance due to insufficient inter-agent coordination.

To address the issue of insufficient cooperation in DTDE,  communication-based knowledge sharing has been proposed, where agents exchange local observations or processed information to improve policy \cite{sukhbaatar2016learning, ding2020learning, kim2020communication}. However, these schemes often exhibit limitations in interpretability and generalizability. Action advising, inspired by teacher-student paradigms,  significantly reduces both training and decision-making costs by having teachers directly suggest actions rather than share raw information with better interpretability \cite{zhu2021q, guo2023explainable, cons}. Yet, current implementations face challenges: students may blindly follow advice from poorly chosen teachers, slowing training or degrading performance, and over-imitation can limit cooperation and exploration \cite{zhu2021q, cons}.

We claim that the key problem causing the above issues is the advice adoption scheme. Agents are distributed and observe the environment locally and partially. Even if a teacher agent is highly experienced in its field, its opinion may be useless for an agent facing a totally different scenario. Only agents sharing a common perceptual consensus can provide valuable action suggestions to others, since they are facing similar scenarios. Inspired by this insight and sociological principles of consensus \cite{snyman2020consensus}, we propose a \textit{Consensus-based Communication and Knowledge Sharing} (CCKS) framework. In CCKS, local observations are regarded as multiple views of the global state, with consensus representing the universal information extracted from these views. 

The key contributions include: (1) pioneering consensus learning to assist action advising, enabling faster convergence to shared environmental understanding; (2) adaptive teacher-student role generation for efficient multi-source knowledge transfer; (3) selective advice adoption based on consensus similarity, ensuring integration of relevant knowledge; and (4) a "think twice before you leap" mechanism where agents resample when low-probability actions are suggested, avoiding premature suboptimal decisions. The framework is plug-and-play, compatible with various DTDE algorithms.

We validate the proposed CCKS framework using two widely recognized benchmark environments: the StarCraft II Multi-Agent Challenge environment \cite{samvelyan2019starcraft} and the Google Research Football environment \cite{kurach2020google}. Integrating CCKS with Independent Q-Learning (IQL) \cite{tampuu2017multiagent} accelerates learning, improves decision accuracy, and enhances stability. Ablation studies confirm the benefits of consensus learning and the "think twice" mechanism, demonstrating CCKS's effectiveness and potential for broader MARL applications.

% % \renewcommand{\dblfloatpagefraction}{.95}
%     \begin{figure*}[tbp]
%     % \hspace{-10mm}
%     \centering
%     	% \caption{Geographical location and relationship of four types of bike stations }
% 		\includegraphics[width=0.72\linewidth]{figure/f5_21.pdf} 
%   \label{Fig.1(a)}
% \vspace{-0.1cm}
% 	\caption{The first part of an overview of CCKS. This part illustrates the communication process among agents, which primarily encompasses the Observation Request Broadcast, Teacher Strategy Response, and the Shared Knowledge Learning segment within the Consensus-based Knowledge Understanding phase. Specifically, the blue sections indicate the agents' Observation Request Broadcast, the red sections signify the Teacher Strategy Response, the green sections represent the agents' Shared Knowledge Learning.}
%       \label{Fig.1-1}
% 	\end {figure*}
% % \vspace{-0.5cm}

\section{RELATED WORK}
\subsection{Distributed execution training framework}
Distributed execution training frameworks mainly include centralized training distributed execution (CTDE) \cite{lowe2017multi, rashid2020monotonic, iqbal2019actor, liu2020pic} and distributed training distributed execution (DTDE) \cite{tan1993multi, tampuu2017multiagent, peng2021facmac, jiang2022i2q}. CTDE leverages global information during training but executes with local observations. However, it is impractical when inter-agent observation sharing is difficult (e.g., human-computer tasks \cite{huang2015adaptive}) or communication is restricted \cite{ying2005multi}. DTDE, in contrast, relies solely on local observations throughout, offering scalability and robustness. While this may lead to suboptimal policies due to the absence of global information, its scalability makes it promising for real-world MARL applications.

\subsection{Knowledge Sharing framework}
Knowledge sharing in Multi-Agent Reinforcement Learning (MARL) enhances coordination by enabling agents to exchange information via communication \cite{ding2020learning, cons, da2017simultaneously} or centralized training \cite{kim2019learning, gupta2023hammer}. Drawing on Theory of Mind (ToM) \cite{tomasello2005understanding}, agents leverage others' insights to inform action selection, mirroring human collaboration. We categorize schemes into observation-level and policy-level sharing. Observation-level approaches (e.g., \cite{sukhbaatar2016learning, ding2020learning}) share local observations or embeddings to enrich perception. Policy-level methods convey higher-level information, such as predicting future intentions \cite{kim2020communication} or assigning teacher-student roles \cite{kim2019learning, cons} for novices to follow experts without independent decisions.

\subsection{Contrastive Learning}
Contrastive Learning has become a prominent paradigm in MARL, particularly for enhancing decentralized policies. It helps agents distinguish between similar and dissimilar experiences, which is essential for learning effective cooperative strategies. In MARL, contrastive learning aligns agent representations toward a consensus, improving stability and convergence \cite{consensus0}. For instance, \cite{contrastive1} introduces a method that uses contrastive learning to decompose the value function while accounting for individual and collective agent identities. Another work \cite{contrastive2} presents a framework for learning grounded communication via contrastive objectives.

\section{BACKGROUND}
\subsection{Problem Formulation}
We formalize our cooperative multi-agent reinforcement learning as a partially observed Markov game for $n$ agents \cite{littman1994markov}, which can be defined by a tuple $(\mathcal{N}, \mathcal{S}, \mathcal{O}, \mathcal{A}, \Omega, \mathcal{P},\left\{\mathcal{R}^i\right\}_{i \in \mathcal{N}}, \gamma)$. Here, $\mathcal{N}=\{1,...,n\}$ denotes the set of $n$ agents; $\mathcal{S}$ is the state space; $\mathcal{O} = O^1 \times ... \times O^n$ denotes the joint observation space; $\mathcal{A} = A^1 \times ... \times A^n$ denotes the joint action space. For each agent $i \in \mathcal{N}$ at each time step, it can only obtain its local observation $o_i \in O^i$ from the observation function $\Omega(s,i)$ where $s \in S$, and then choose the action $a^i \in A^i$ according to the policy $\pi_{\omega_i}(a^i|o^i): O^i \to A^i$, where $\omega_i$ parameterizes the policy. The environment changes according to a state transition function $\mathcal{P}: \mathcal{S} \times \mathcal{A} \to \mathcal{S}'$, which returns a probability distribution of a new state $\mathcal{S'}$ based on the current state $\mathcal{S}$ and the joint action $\mathcal{A}$. Similarly, the agent can obtain new observations $o_i'$ through the observation function $o_i'=\Omega(s',i)$. After that, each agent $i \in \mathcal{N}$ will receive its own reward $r^i_t$ at the current time step $t$ based on the reward function $\mathcal{R}^i : \mathcal{S} \times \mathcal{A} \times \mathcal{S}' \to \mathcal{R}$. The purpose of each agent is to optimize the parameter $\omega_i$ of the policy $\pi_{\omega_i}$ to maximize the expected discounted reward $\mathcal{E}[\sum_{t=0}^T \gamma^t r_t^i \mid \pi_{\omega_i}]$, where $\gamma$ is the discount factor and $T$ is the total time step.

To enhance coordination, we integrate a communication mechanism where each agent \( i \) can send and receive messages \( m_i \in \mathcal{M} \) based on its local observation \( o_i \) and action \( a^i \). The communication strategy \( \phi_{\omega_i^c}(m^i|o^i, a^i): O^i \times A^i \to \mathcal{M} \) determines the message content. Received messages are used to update each agent's policy \( \pi_{\omega_i} \), enabling more informed and coordinated actions. This mechanism improves collective performance by allowing agents to share critical information and adjust their strategies accordingly.

\begin{figure*}[tbp]
    \centering
    \subfigure[]{
        \includegraphics[width=0.49\linewidth]{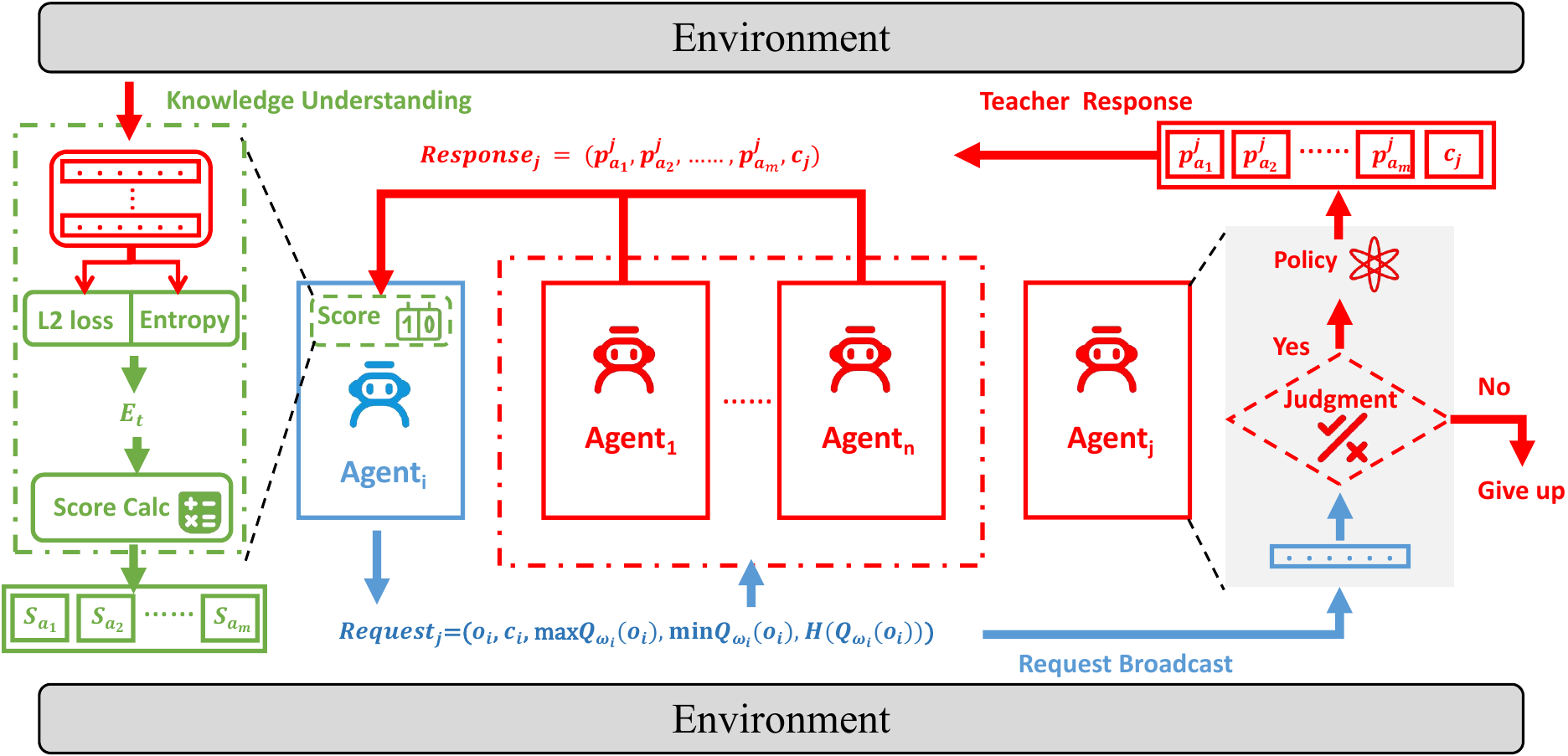} \label{Fig.1-1}
    }
    \subfigure[]{
        \includegraphics[width=0.47\linewidth]{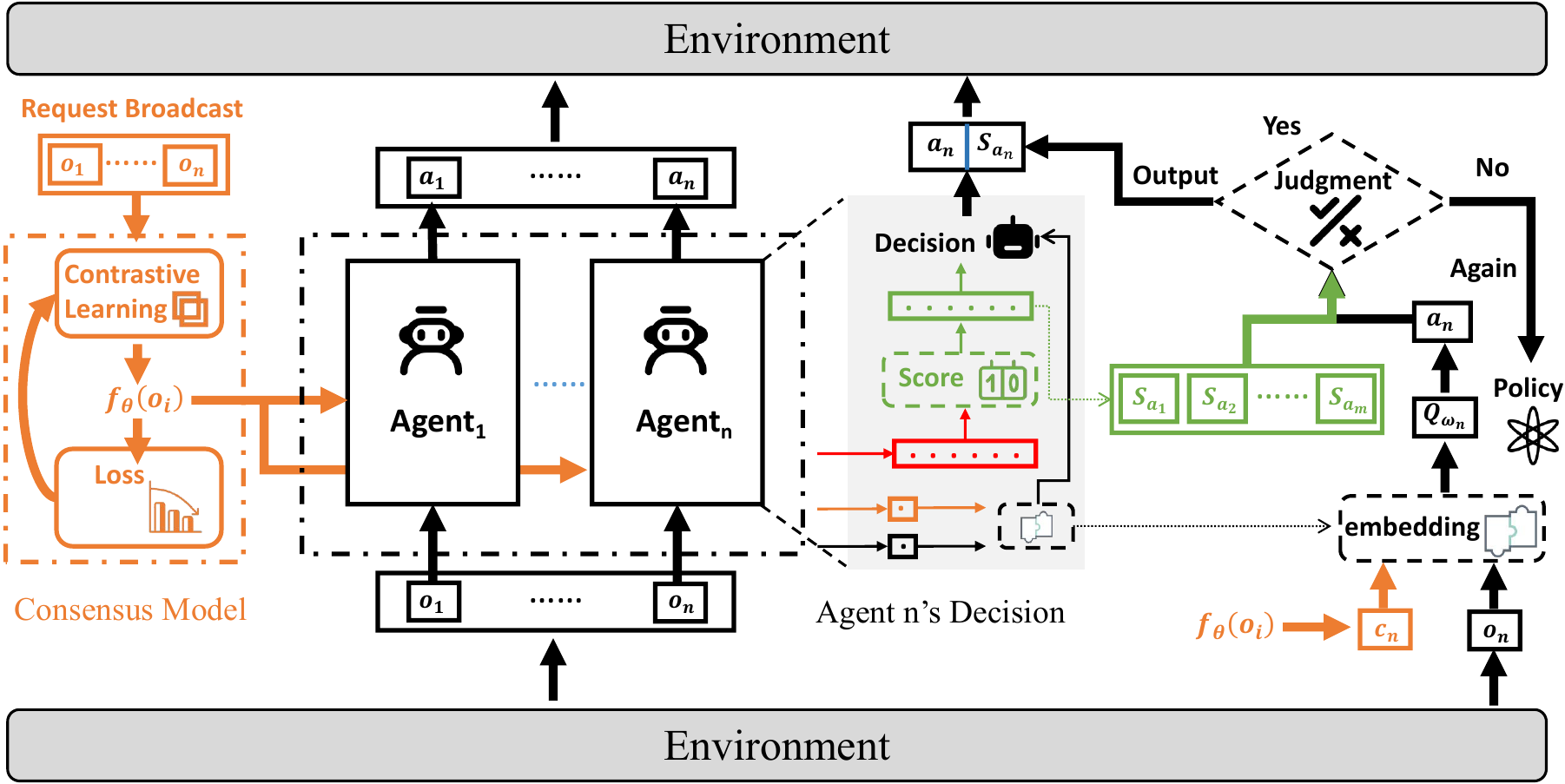} \label{Fig.1-2}
    }
    \vspace{-0.1cm}
    \caption{This figure provides an overview of CCKS. (a) The communication process among agents. Specifically, the blue sections indicate the agents' Observation Request Broadcast, the red sections signify the Teacher Strategy Response, and the green sections represent the agents' Shared Knowledge Learning. (b) depicts the interaction process between the agents and the environment. Specifically, the orange sections denote Consensus Model Training, and the black sections correspond to Actions Sampling.}
    \label{Fig4}
\end{figure*}

\section{METHOD}

% In this section, we introduce CCKS, which consists of four key stages: Consensus Model Training, Observation Request Broadcast, Teacher Strategy Response, and Consensus-based Knowledge Understanding. In the first stage, we employ contrastive learning to train a consensus model that helps agents quickly form a shared understanding of the global state, reducing learning complexity and enhancing coordination. In the second stage, each agent independently broadcasts its observation request to ensure timely information sharing and address the issue of information asymmetry among agents. In the third stage, agents use a Q-function and information entropy to estimate the experience level of the requesting agent, dynamically deciding whether to assume the role of a teacher, thereby addressing the issue of fixed teacher roles in existing methods and enhancing the flexibility of knowledge sharing. In the last stage, agents use information entropy and the consensus mechanism to determine the degree and weight of knowledge sharing, ensuring that only relevant and high-quality information is absorbed, preventing the propagation of suboptimal strategies, and improving the quality of knowledge transfer. These stages collectively enhance coordination and learning efficiency among agents, with the detailed training process described in Figures \ref{Fig.1-1} and \ref{Fig.1-2} and the algorithm can be found at the extended version\cite{arxiv}.

In this section, we introduce CCKS's four stages: Consensus Model Training, Observation Request Broadcast, Teacher Strategy Response, and Consensus-based Knowledge Understanding. First, contrastive learning trains a consensus model for shared environmental understanding, reducing complexity. Second, agents broadcast observation requests for timely sharing and information symmetry. Third, agents dynamically assume teacher roles using Q-function and entropy, enhancing flexibility. Finally, agents filter advice via entropy and consensus, ensuring quality knowledge transfer. See Figures \ref{Fig.1-1} and \ref{Fig.1-2}.
% and algorithm in the extended version \cite{arxiv}.

\subsection{Consensus Model Training}
In the conventional action advising scheme, the agent often learns only from the individual experiences of more capable agents, rather than learning to achieve better cooperative outcomes, which is the core problem we aim to solve. In CCKS, we introduce a consensus learning model $f_{\theta}$ into the DTDE paradigm. Each agent uses its local observation $o_i$ to derive a consensus representation $c_i$ of the global state, which supplements $o_i$ for decision-making. The consensus inferred by the agents serves as a shared signal, akin to a global state, guiding their actions toward a common goal. The specific formula of the consensus \( c_i \) is as follows:

% This consensus \( c_i \) encapsulates the agent's interpretation of the state, incorporating its unique perspective and learned features, which is crucial for aligning the agents' understanding of the environment. 
\begin{equation} c_i = f_{\theta}(o_i)\label{eq3}
\end{equation}
We use a three-layer multilayer perceptron (MLP) with ReLU activation to train the consensus model via contrastive learning. The ultimate goal of the consensus model is to ensure that the consensus outputs of all agents align as closely as possible. 

% This consensus is then utilized as an explicit input to the policy network and as an important part of calculating teacher weights according to Equation \ref{eq8} and Equation \ref{eq11}.

\subsection{Observation Request Broadcast}

In the MARL framework, the Observation Request Broadcast serves as a pivotal mechanism for facilitating inter-agent communication and knowledge dissemination. This phase is essential for enabling each agent to share its current state representation and associated Q-value estimations with its counterparts, thereby fostering a collaborative learning environment.

Each agent \( i \) within the system computes a consensus \( c_i \) for its observed state \( o_i \) using its consensus model. Once the consensus \( c_i \) is computed, agent \( i \) proceeds to broadcast a request to all other agents. This request comprises the following components:
\begin{equation}
Request_i = \left( o_i, c_i, \max Q_{\omega_i}(o_i), \min Q_{\omega_i}(o_i), H(Q_{\omega_i}(o_i))  \right)\label{eq4}
\end{equation}
where \( H(\cdot) \) represents the entropy of the Q-values, while \( \max Q_{\omega_i}(o_i) \) and \( \min Q_{\omega_i}(o_i) \) denote the maximum and minimum Q-value estimations for state \( o_i \), respectively. 
% These Q-values provide insights into the agent's value assessments of the current state, which are vital for informed decision-making and strategy alignment among the agents.

\subsection{Teacher Strategy Response}

In the Teacher Strategy Response phase each agent \( j \) will decide whether to share its knowledge with others to facilitate knowledge transfer and collaboration.

Each agent \( j \) evaluates the received observations and Q-value estimations from its peers to decide on knowledge sharing. Agent \( j \) shares its knowledge with agent \( i \) only if both of the following conditions are met:
\begin{enumerate}
    \item The observation \( o_i \) is not new to agent \( j \) (i.e., \( n_{o_i}^j \ge 1 \)) and the entropy of agent \( j \)'s Q-values for \( o_i \) is not greater than that of agent \( i \):
    \begin{equation}
    n_{o_i}^j \ge 1 \quad \text{and} \quad H(Q_{\omega_i}(o_i)) \ge H(Q_{\omega_j}(o_i)) \label{eq5}
    \end{equation}
    \item The Q-value estimation of agent \( j \) for state \( o_i \) lies within the interval defined by agent \( i \)'s min and max Q-values:
    \begin{equation}
    \min Q_{\omega_i}(o_i) \le Q_{\omega_j}(o_i) \le \max Q_{\omega_i}(o_i) \label{eq6}
    \end{equation}
\end{enumerate}
If either condition fails, agent \( j \) refrains from sharing knowledge with agent \( i \). Otherwise, it proceeds to share its action probabilities \( p_{a_l}^j \) for each action \( a_l \) in the action space, computed using its policy network \( Q_{\omega_j} \):
\begin{equation}
Response_j = \left(p_{a_1}^j, p_{a_2}^j, \ldots, p_{a_m}^j, c_j \right)\label{eq7}
\end{equation}

\subsection{Consensus-based Knowledge Understanding}

\subsubsection{Shared Knowledge learning}

In the Consensus-based Knowledge Sharing phase, each agent \( i \) receives responses from its teachers who denoted as \( T \) and computes a guiding coefficient \( E_t \) for each teacher \( t\in T \) based on the discrepancy between its own consensus \( c_i \) and that of teacher \( t \), as well as the entropy of the teacher's Q-values:
\begin{equation}
E_t = -\left\| c_i - f_{\theta}(o_t) \right\| \cdot H(Q_{\omega_t}(o_i))\label{eq8}
\end{equation}
Here, the consensus difference \( \|c_i - f_{\theta}(o_t)\| \) measures agent-teacher alignment, while \( H(Q_{\omega_t}(o_i)) \) denotes Q-value entropy, reflecting teacher uncertainty. Guiding weights \( \mu_t \) are then computed via softmax normalization:

% When the entropy is equal to 0, it means that the teacher agent thinks that an action must be taken in this situation; conversely, the greater the entropy, the higher the uncertainty of the teacher agent about the current situation. 
\begin{equation}
\mu_t = \frac{e^{E_t}}{\sum_{t} e^{E_t}}\label{eq9}
\end{equation}
These weights determine the influence of each teacher on the agent \( i \), ensuring that more reliable and diverse knowledge sources have a greater impact. With these weights, each action's score \( S_{a_l} \) is calculated as:
\begin{equation}
S_{a_l} = \sum_{t} \mu_t \cdot (p_{a_l}^t) - S_{\text{mean}} \label{eq10}
\end{equation}
where \( S_{\text{mean}} = \frac{1}{m}\sum_{l=1}^{m} \sum_{t} \mu_t \cdot p_{a_l}^t \) is the average score across all actions. This scoring mechanism uses the probability of the teachers' policies to score each action, guiding the agent to adjust its policy toward more optimal actions.

\subsubsection{Actions sampling}
In the conventional action advising scheme, the actions of the student agent are often directly overwritten or modified, which can conflict with its policy, impair learning, and misalign with its mental model. Hence, our scheme avoids directly influencing the agent's decision-making process. In the decision-making process of the agent with a Q-function, we replace the regular observation \(o_i\) with the concatenation of observation \(o_i\) and consensus \(c_i\); the action of agent \(i\) is then selected as follows:
\begin{equation}
a_i = \text{Sample}\big(Q_i(\text{embedding}(o_i,c_i), a_i)\big)\label{eq11}
\end{equation}

Meanwhile, the scoring mechanism participates in the process through two methods, namely ``Think twice before you leap`` and reward shaping. Given the hyper-parameter \( K \), if an agent selects an action from the set of \( K \) actions with the lowest scores (deemed by teachers as likely to ``fall into pitfalls'''), it will perform a secondary sampling, inspired by the human tendency to ``think twice''. Additionally, considering that ``Think twice before you leap'' does not effectively utilize high-scoring actions, we use the scores for reward shaping. Specifically, the original reward \(r\) is updated using the hyper-parameter \( \tau \) (\(\tau \in [0,1] \)) as follows:
\begin{equation}
r_{\text{new}} = r + \tau \cdot S_{a_i}\label{eq12}
\end{equation}

\begin{figure}[tbp]
    \centering
    	% \caption{Geographical location and relationship of four types of bike stations }
	\subfigure[]{
		\includegraphics[width=0.45\linewidth]{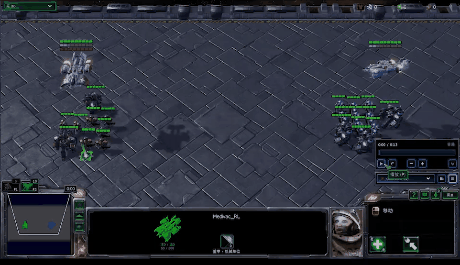} 
  \label{Fig.3(a)}
	}
	% \hspace{5mm}
 \centering
	\subfigure[]{
		\includegraphics[width=0.45\linewidth]{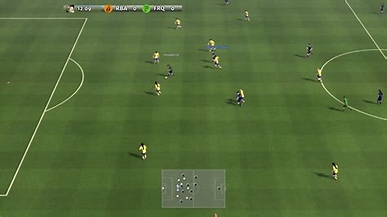}
  \label{Fig.3(b)}
	}	
\vspace{-0.1cm}
	\caption{Experiment environments. (a) The StarCraft II Multi-Agent Challenge (SMAC). (b) Google Research Football (GRF).}
    \label{Fig.en}
	\end {figure}
% \vspace{-0.3cm}

 \section{EXPERIMENTS}

In this section, we evaluate the performance of CCKS as a new DTDE framework by comparing it with baseline algorithms. In our experiments, Independent Q-Learning (IQL) \cite{tampuu2017multiagent} was used as the base algorithm to assess the effectiveness of CCKS. We implement CCKS and compare it with different baselines, including IQL \cite{tampuu2017multiagent}, CONS \cite{cons}, I2Q \cite{jiang2022i2q}, and AdHocTD \cite{da2017simultaneously}. Finally, we study the conditions and components that affect CCKS's performance through ablation experiments. 

Our testing environments include the StarCraft Multi-Agent Challenge (SMAC) \cite{samvelyan2019starcraft} and Google Research Football (GRF) \cite{kurach2020google}, depicted in Figure \ref{Fig.en}. SMAC provides real-time strategy micro-management tasks in StarCraft II. GRF simulates football matches requiring coordinated tactical execution. For each environment, we conduct experiments on multiple maps and scenarios. All results are reported as the mean and standard deviation over five random seeds.

\begin{figure*}[htbp]
    \centering
    \subfigure[]{
        \includegraphics[width=0.23\linewidth]{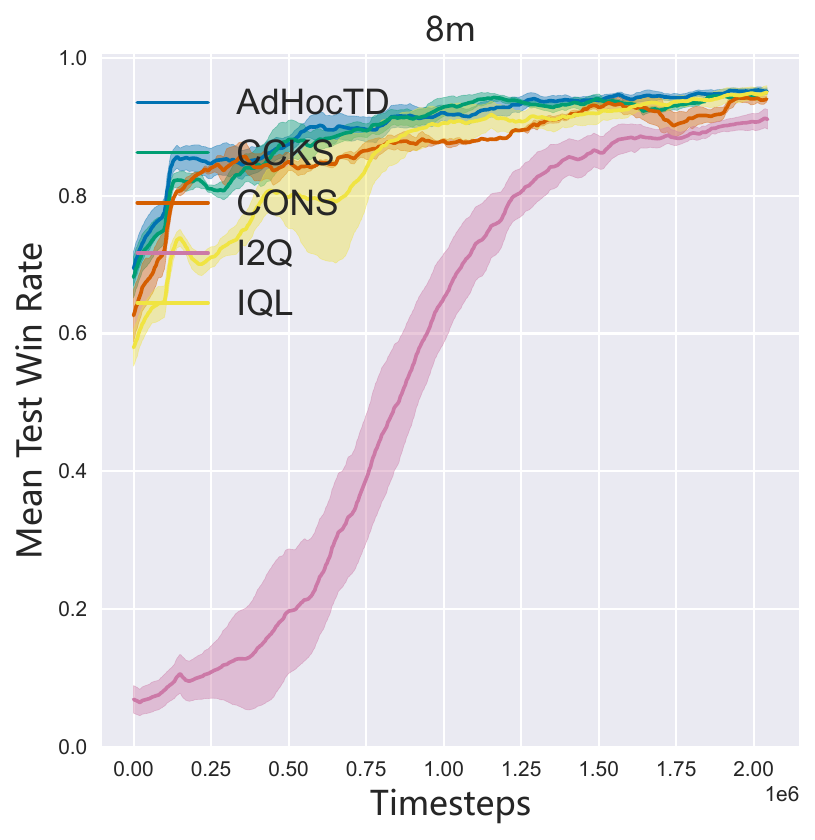} \label{Fig.4(a)}
    }
    \subfigure[]{
        \includegraphics[width=0.22\linewidth]{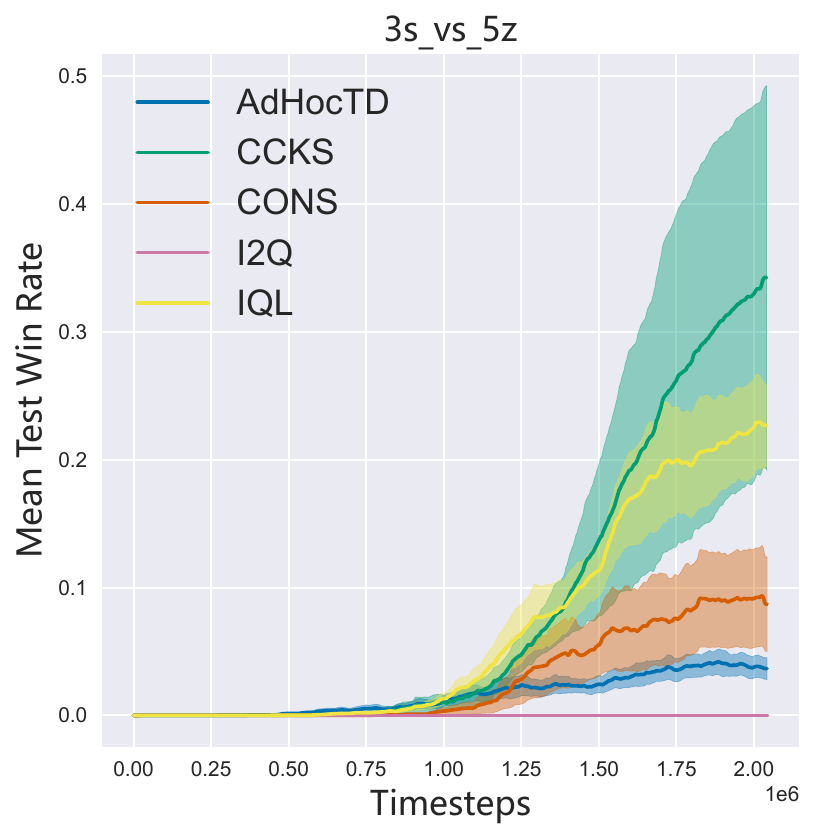} \label{Fig.4(b)}
    }
    \subfigure[]{
        \includegraphics[width=0.22\linewidth]{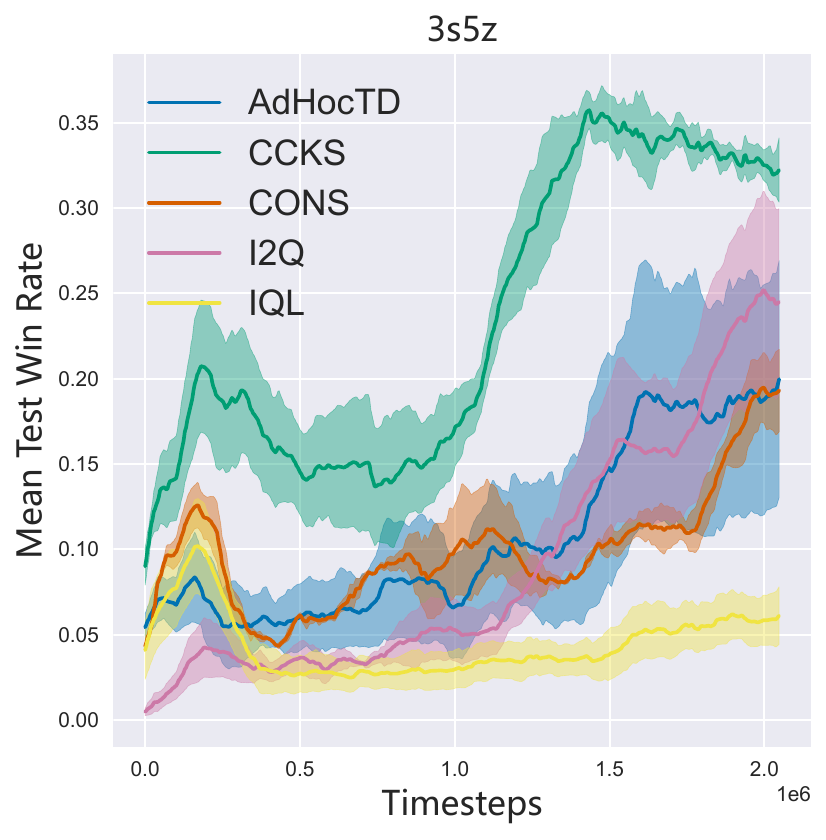} \label{Fig.4(c)}
    }
    \subfigure[]{
        \includegraphics[width=0.22\linewidth]{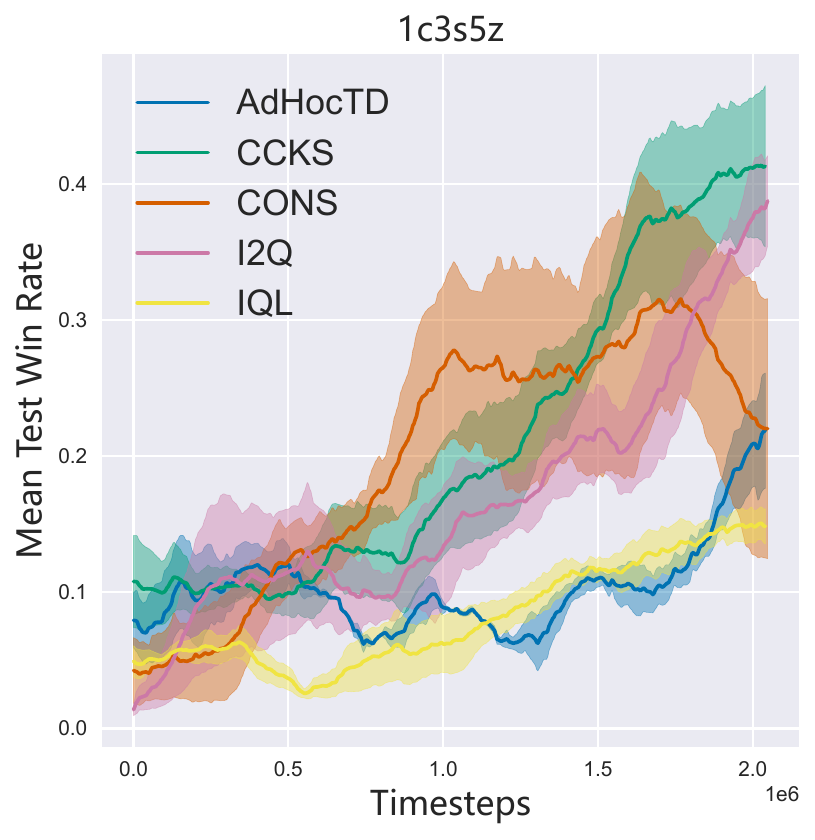} \label{Fig.4(d)}
    }
    \vspace{-0.1cm}
    \caption{Experimental results on the StarCraft Multi-Agent Challenge Environment: The mean test win rate of the five algorithms over 5 seeds under a partially observable SMAC task.}
    \label{Fig4}
\end{figure*}

\begin{figure*}[htbp]
 \centering
	\subfigure[]{
		\includegraphics[width=0.22\linewidth]{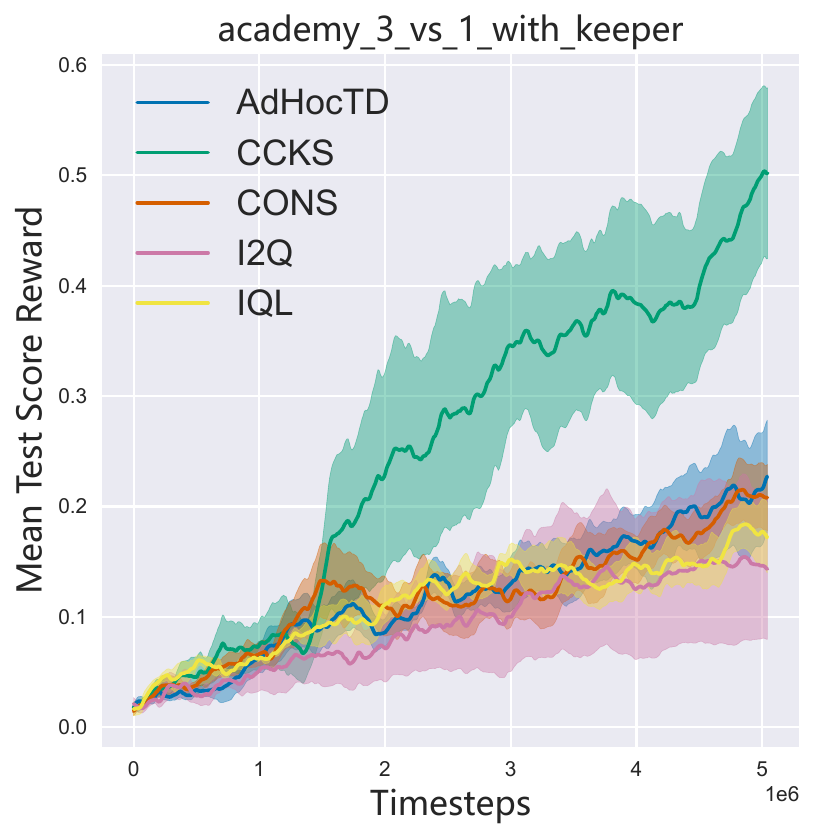} \label{Fig.5(a)}
	}
	% \hspace{2mm}
 \centering
	\subfigure[]{
		\includegraphics[width=0.22\linewidth]{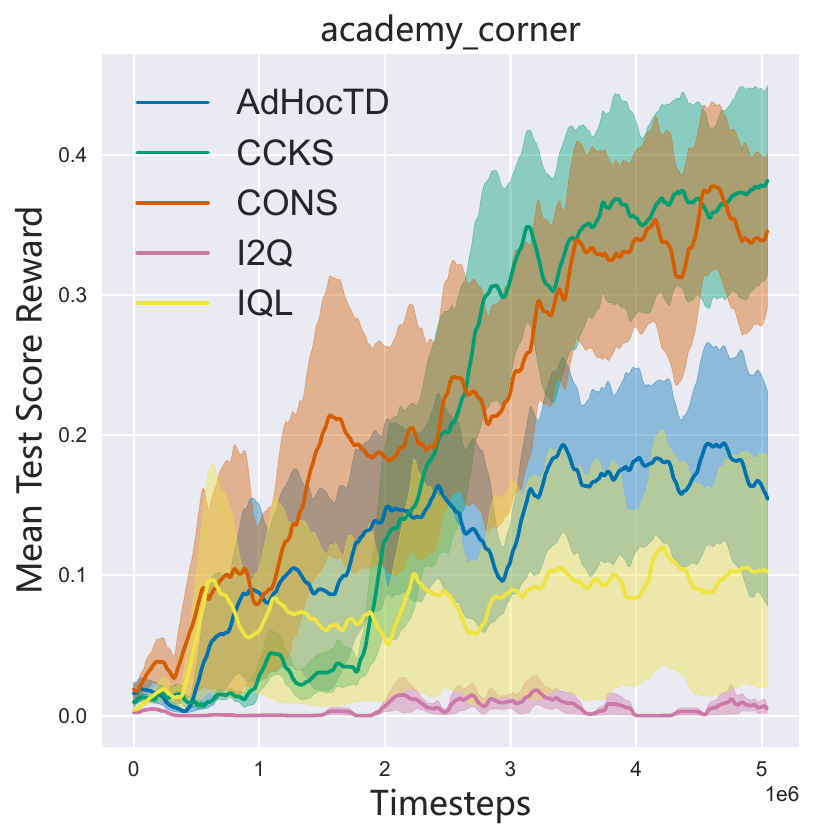} \label{Fig.5(b)}
	}	
     \centering
	\subfigure[]{
		\includegraphics[width=0.22\linewidth]{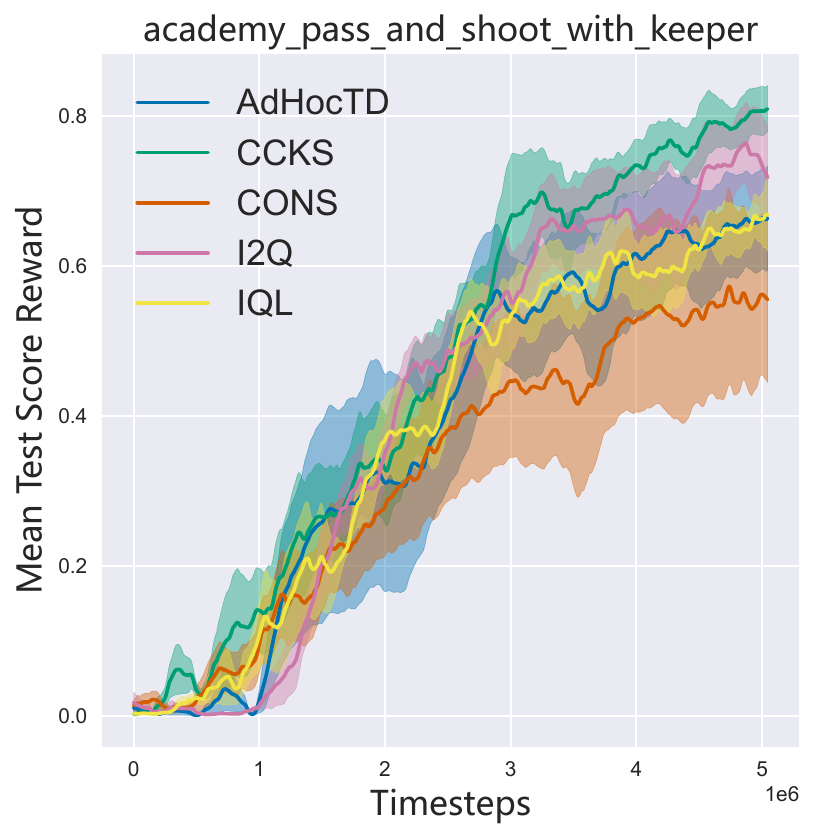} \label{Fig.5(c)}
	}	
     \centering
	\subfigure[]{
		\includegraphics[width=0.22\linewidth]{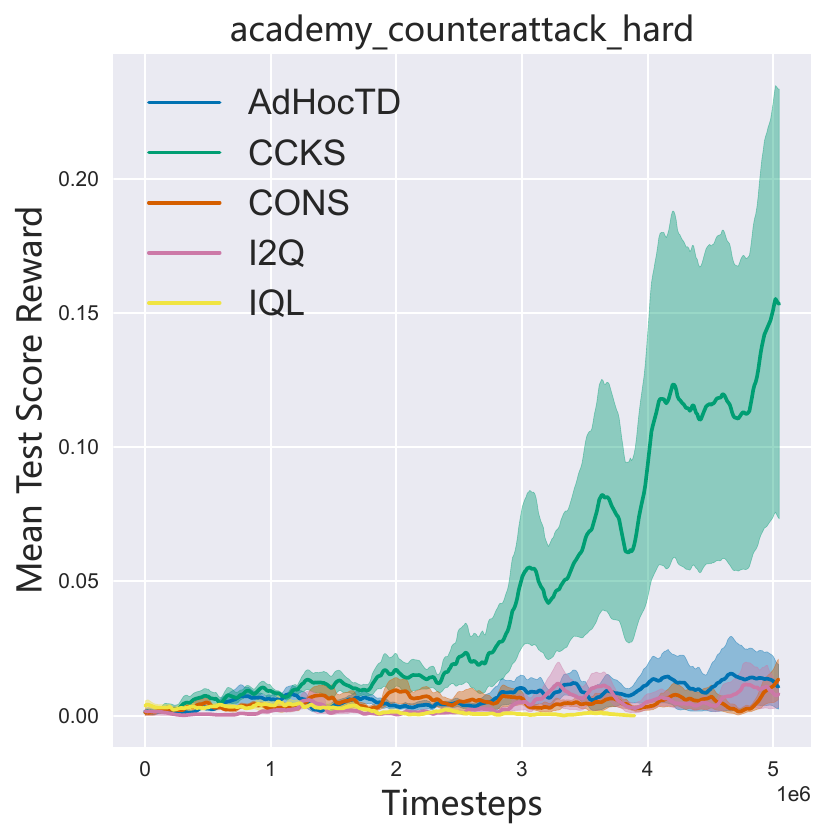} \label{Fig.5(d)}
	}	
    \vspace{-0.1cm}
  \caption{Experimental results on the Google Research Football Environment: The mean test score reward of the five algorithms over 5 seeds under a partially observable GRF task.}\label{Fig5}
\end{figure*}

\subsection{Performance}
Figure \ref{Fig4} presents the StarCraft II results, showing the mean test win rate over timesteps, with shaded areas indicating the error in performance across different seeds. The learning speed and overall performance of the CCKS algorithm have been significantly enhanced. The extent of improvement varies across different scenarios. Notably, during the initial stages of training, CCKS exhibits a substantial performance gap compared to other algorithms such as IQL and I2Q. We attribute this to the consensus mechanism of CCKS, which enables implicit cooperation among agents. In contrast, agents in the IQL algorithm must independently explore the environment, often selecting actions that are more advantageous for individual agents rather than the team. For AdHocTD, excessive imitation of the teacher agent frequently leads to suboptimal outcomes. The I2Q algorithm trains an ideal transition model using global information; however, the complexity of this information results in slower performance improvements. CONS provides richer information for agent decision-making through action suggestions, but the limited knowledge sharing for a single action leads to insufficient learning. Compared with these methods, CCKS establishes consensus under communication conditions and implicitly learns knowledge through a scoring mechanism. This approach enables agents to ``think twice'' before acting, thereby achieving better overall performance.

Figure \ref{Fig5} presents the experimental outcomes from the Google Football environment, illustrating the performance of various algorithms over timesteps, focusing on mean test episode length and mean test score reward. The results indicate that CCKS is the most effective algorithm, achieving both shorter episode lengths and higher score rewards over time, suggesting that it is better at learning efficient and high-quality actions in a multi-agent setting compared to the other algorithms tested, such as AdHocTD, CONS, I2Q, and IQL, which exhibit more instability and less consistent performance improvements.

% Figure \ref{Fig5} presents the experimental outcomes from the Google Football environment, illustrating the performance of various multi-agent reinforcement learning algorithms over a series of timesteps, focusing on mean test episode length and mean test score reward. The results indicate that CCKS is the most effective algorithm, achieving both shorter episode lengths and higher score rewards over time, suggesting that it is better at learning efficient and high-quality actions in a multi-agent setting compared to the other algorithms tested, such as AdHocTD, CONS, I2Q, and IQL, which exhibit more instability and less consistent performance improvements.

\begin{figure*}[htb]
 \centering
	\subfigure[]{
		\includegraphics[width=0.22\linewidth]{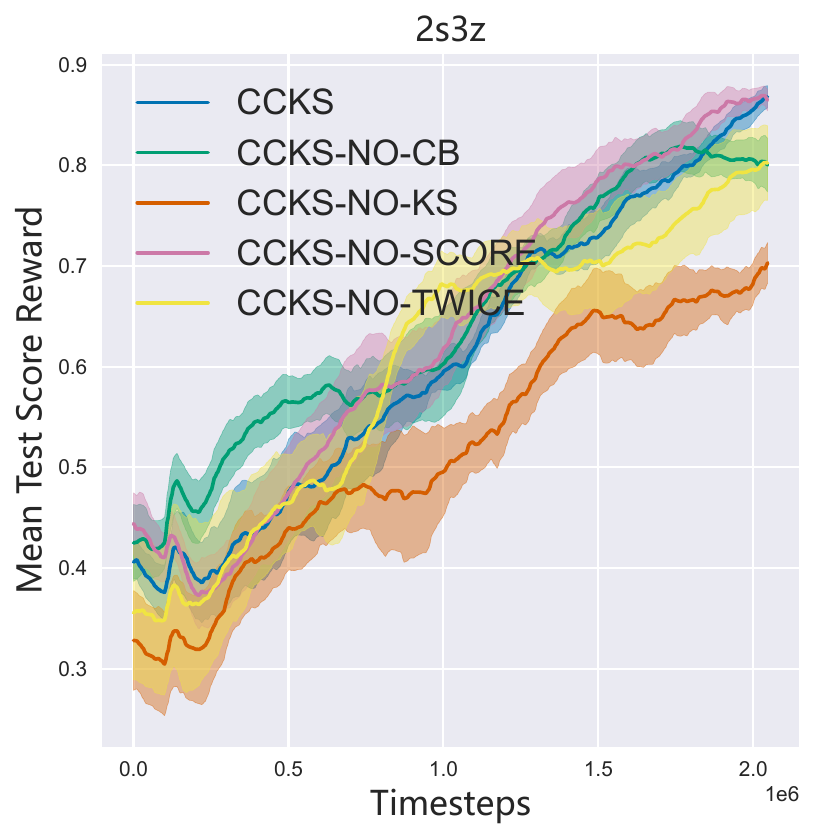} \label{Fig.6(a)}
	}
	% \hspace{2mm}
 \centering
	\subfigure[]{
		\includegraphics[width=0.22\linewidth]{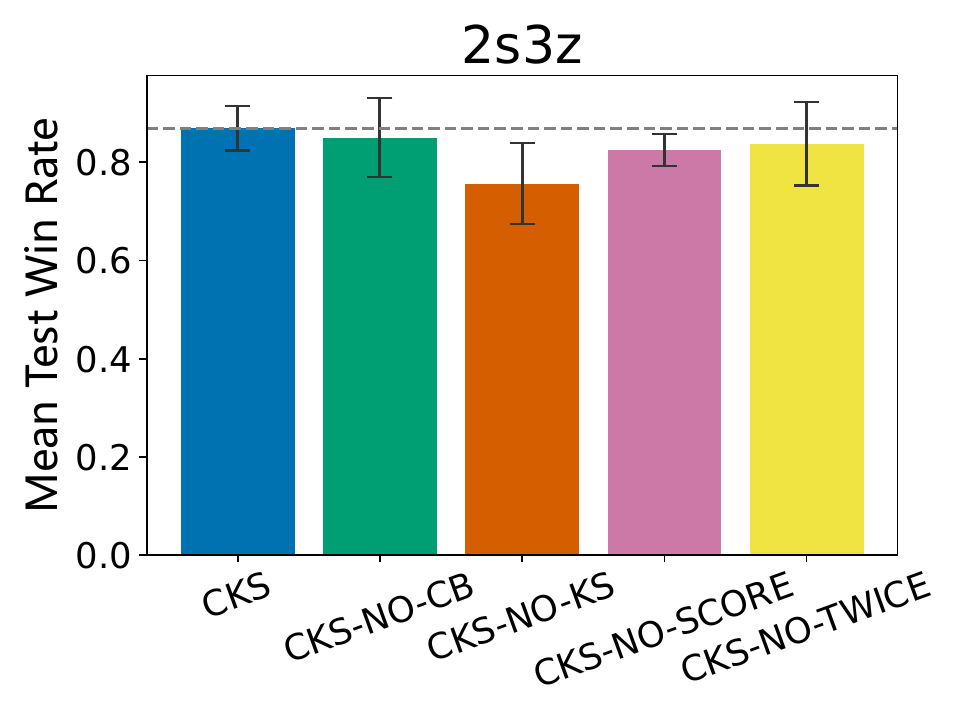} \label{Fig.6(b)}
	}	
 \centering
	\subfigure[]{
		\includegraphics[width=0.22\linewidth]{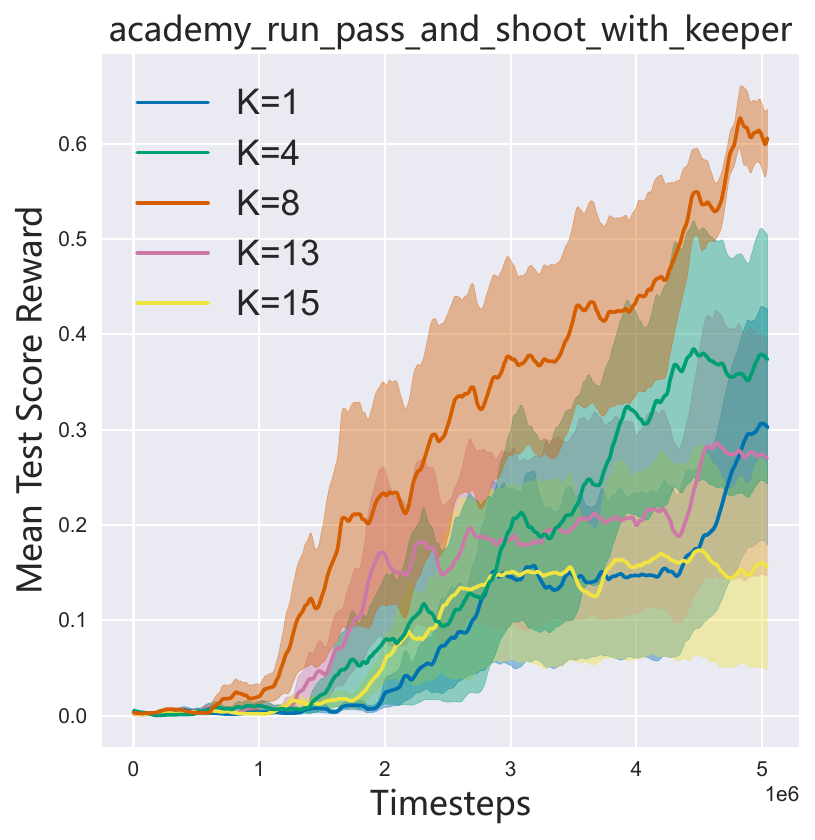} \label{Fig.6(c)}
	}
	% \hspace{2mm}
 \centering
	\subfigure[]{
		\includegraphics[width=0.22\linewidth]{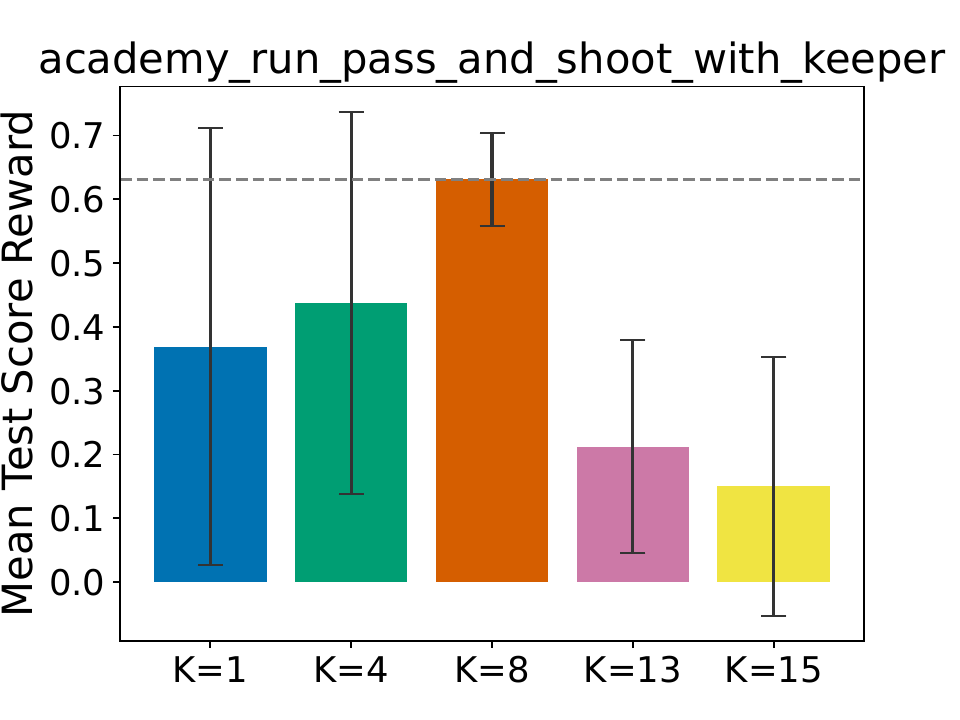} \label{Fig.6(d)}
	}
    \vspace{-0.1cm}
  \caption{Ablation experimental results on SMAC and GRF showing the influence of the components and \( K \) of CCKS.}\label{Fig6}
\end{figure*}

\subsection{Further Analysis}
\subsubsection{Ablations}
To assess the efficacy of the CCKS algorithm, an ablation study was conducted on the \textit{2s3z} map within the SMAC environment, evaluating the contributions of its key components: the consensus model, knowledge sharing, scoring mechanism, and the ``think twice'' mechanism. Configurations lacking these features were denoted as \text{CCKS-NO-CB}, \text{CCKS-NO-KS}, \text{CCKS-NO-SCORE}, and \text{CCKS-NO-TWICE}. Results, as shown in Figure \ref{Fig.6(a)} and Figure \ref{Fig.6(b)}, indicate that CCKS achieves the highest mean test win rate, significantly outperforming other configurations as the number of timesteps increases. This suggests that the full CCKS algorithm is most effective in learning strategies leading to higher win rates. \text{CCKS-NO-CB} and \text{CCKS-NO-KS} show improvements in win rate over time but perform less consistently and slightly lower compared to the full CCKS algorithm. \text{CCKS-NO-SCORE} and \text{CCKS-NO-TWICE} exhibit more variability, with CCKS-NO-SCORE showing a particularly noticeable decline in win rate at certain points, indicating that the absence of scoring mechanisms or the ``think twice'' mechanism may negatively impact learning and overall performance. The ablation studies suggest that each component of the CCKS algorithm contributes to its overall performance, and removing any of these components leads to a decrease in effectiveness.

Additionally, the impact of the hyperparameter \( K \) on CCKS performance was investigated within the GRF environment. With all other settings constant, performance under various \( K \) values was compared. Figure \ref{Fig.6(c)} and Figure \ref{Fig.6(d)} illustrate the training curves and execution outcomes. In the GRF scenario, smaller \( K \) values result in slower learning speeds and significant variance. As \( K \) increases, performance improves significantly compared to smaller \( K \) values. However, as \( K \) continues to increase, despite faster initial training speeds, final performance declines. This suggests that an appropriate increase in \( K \) can enhance learning speed, but an excessively large \( K \) can decrease exploratory capabilities, leading to local optima. Selecting an appropriate \( K \) value can significantly enhance CCKS's experimental outcomes.

\subsubsection{Limitations and future work}
A limitation of CCKS is the added computational overhead from the consensus model during training—a trade-off for improved learning efficiency. This overhead must be considered in resource-limited or real-time settings. Moreover, since knowledge sharing relies on probabilities over all discrete actions, expanding the action space significantly increases communication overhead, slowing training. This is especially pertinent as real-world tasks often have larger action spaces than those in GRF or SMAC.

Future work should also explore the further alignment of the relationships between teacher and student agents, as well as the design of reward mechanisms. Employing interpretable learning processes can foster policy models closer to human cognition, improving AI interactions.

\section{CONCLUSIONS}

In this paper, we propose the CCKS framework as a plug-and-play component to enhance the capabilities of existing cooperative MARL algorithms in order to address the lack of cooperation in the DTDE paradigm. CCKS introduces consensus learning, which takes each agent's local observation as input and outputs a consensus representation as auxiliary information. A knowledge sharing process enables agents to obtain guidance from others, where student agents compute weights for teachers and score actions based on information entropy and L2 loss. Using the resulting score table, the ``Think twice before you leap'' mechanism and reward shaping refine agents' actions and rewards. Extensive experiments on SMAC and GRF demonstrate that CCKS significantly improves the performance of baseline cooperative MARL algorithms.

\addtolength{\textheight}{-12cm}   % This command serves to balance the column lengths

%%%%%%%%%%%%%%%%%%%%%%%%%%%%%%%%%%%%%%%%%%%%%%%%%%%%%%%%%%%%%%%%%%%%%%%%%%%%%%%%

%%%%%%%%%%%%%%%%%%%%%%%%%%%%%%%%%%%%%%%%%%%%%%%%%%%%%%%%%%%%%%%%%%%%%%%%%%%%%%%%



% \section*{ACKNOWLEDGMENT}

% The preferred spelling of the word ÒacknowledgmentÓ in America is without an ÒeÓ after the ÒgÓ. Avoid the stilted expression, ÒOne of us (R. B. G.) thanks . . .Ó  Instead, try ÒR. B. G. thanksÓ. Put sponsor acknowledgments in the unnumbered footnote on the first page.

%%%%%%%%%%%%%%%%%%%%%%%%%%%%%%%%%%%%%%%%%%%%%%%%%%%%%%%%%%%%%%%%%%%%%%%%%%%%%%%%

% References are important to the reader; therefore, each citation must be complete and correct. If at all possible, references should be commonly available publications.

\bibliographystyle{IEEEtran}
\bibliography{IEEEabrv,IEEEtran}

\end{document}